\documentclass[ reprint, 
 aps,
 pra,
 showkeys,
 nofootinbib
]{revtex4-1}
\usepackage{amsfonts}
\usepackage{amssymb}
\usepackage{amsmath}
\usepackage{graphicx}
\usepackage{subfigure}
\usepackage{dcolumn}
\usepackage{bm}
\usepackage{setspace}
\usepackage[hang]{footmisc}
\usepackage[sort&compress]{natbib}
\usepackage{multirow}
\usepackage[sort&compress]{cleveref}
\usepackage{nomencl}
\usepackage[english]{babel}

\crefname{equation}{\unskip}{\unskip}
\crefname{figure}{\unskip}{\unskip}
\crefname{section}{\unskip}{\unskip}
\crefname{subsection}{\unskip}{\unskip}

\begin{document}
\let\ref\cref

\title{General contact mechanics theory for randomly rough surfaces with application to rubber
friction}
\author{M. Scaraggi}
\affiliation{DII, Universit\'a del Salento, 73100, Italy}
\affiliation{PGI, FZ J\"ulich, 52425 J\"ulich, Germany}
\author{B.N.J. Persson}
\affiliation{PGI, FZ J\"ulich, 52425 J\"ulich, Germany}
\affiliation{www.MultiscaleConsulting.com}

\begin{abstract}
We generalize the Persson contact mechanics and rubber friction theory
to the case where both surfaces have surface roughness. 
The solids can be rigid, elastic or viscoelastic, and can be homogeneous or layered.
We calculate the contact area, the viscoelastic contribution to the friction force, and
the average interfacial separation as a function of the sliding speed and the nominal contact pressure.
We illustrate the theory with numerical results for a rubber block sliding on a road
surface. We find that with increasing sliding speed, the influence of the roughness on the rubber block
decreases, and for typical sliding speeds involved in tire dynamics it can be neglected.
\end{abstract}

\maketitle

%%%%%%%%%%%%%% main text %%%%%%%%%%%%%%%%
%\begin{multicols}{2}

\vskip 0.3cm \textbf{1 Introduction}

Practically all surfaces in Nature and technology have surface roughness
over many decades in length scales, often extending from the macroscopic
size of the system down to atomic distances\cite{roughness}. For this reason
the interaction between bodies in relative motion is a very complex topic.
Nevertheless, because of its crucial importance for many important
tribological problems, such as the electric and thermal contact resistance%
\cite{Barber,Persson1,Robbins}, rubber friction\cite%
{Persson,Carbone,Scaraggi}, sealing\cite{Dapp}, frictional heating\cite{last}
and tribo-electrification\cite{tribo1,tribo2}, a large effort has recently
been devoted to the contact mechanics between solids with rough surfaces\cite%
{c1,c2,c3,Robbins,c5,c6,c7,c8,c9}.

Archard\cite{Arch}, and Greenwood and Williamson (GW)\cite{GW}, have
presented pioneering studies of the contact between solids with roughness.
These theories approximate the surface asperities with spherical cups to
which they apply the Hertz contact mechanism. The theory of Archard is for a
rather idealized situation, so most contact mechanics studies have used the
GW theory. However, the GW theory neglects the elastic coupling between the
different contact regions (i.e. the lateral or \textit{horizontal coupling}%
), as well as the vertical coupling between hierarchically distributed
asperities (small bumps on top of bigger bumps on top of bigger bumps, and
so on). For surfaces with roughness on many length scales this approximation
leads to qualitative wrong results both for the contact area and the average
interfacial separation as a function of the applied normal force, and other
properties\cite{Rob,Per,Car,Mus}.

A very different approach was developed by Persson\cite{Persson} in 2001.
This theory includes the elastic coupling and can be applied for arbitrary
large normal forces, e.g., even as the contact approaches full contact,
which is very important for some applications like the leak-rate of seals.
The theory is very flexible and can easily be applied to complex situations,
e.g., layered materials and elastoplastic or viscoelastic materials. The
original theory was developed for a viscoelastic solid with a perfectly flat
surface in contact with a hard, randomly rough, substrate. In this paper we
extend this theory to the case where both solids have surface roughness, and
arbitrary viscoelastic properties, e.g., a viscoelastic solid in contact
with an elastic solid.

\begin{figure}[tbp]
\begin{center}
\includegraphics[
                        width=0.3\textwidth,
                        angle=0]{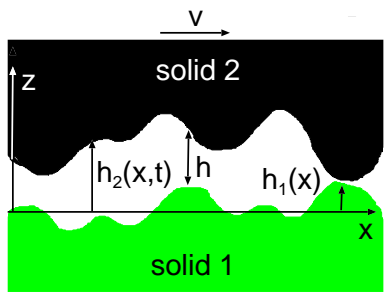}
\end{center}
\caption{Two solids with the surface profiles $h_{1}(\mathbf{x})$ and $h_{2}(%
\mathbf{x},t)$. The gap $h(\mathbf{x})=h_{2}(\mathbf{x},t)-h_{1}(\mathbf{x})$%
. If the upper solid moves at a constant velocity $\mathbf{v}=(v_{\mathrm{x}%
},v_{\mathrm{y}})$ parallel to the bottom solid, we have $h_{2}(\mathbf{x}%
,t)=h_{2}(\mathbf{x}-\mathbf{v}t)$. }
\label{pich1.ps}
\end{figure}

Consider the \textit{frictionless} contact between \textit{elastic} solids
with nominally flat surfaces in normal (quasi-static) approach. Assume that
the solids have roughness described by the height profiles $h_{1}(\mathbf{x}%
) $ and $h_{2}(\mathbf{x})$ [where $\mathbf{x}=(x,y) $ is a two-dimensional
position vector in the surface $xy$-plane], and different elastic properties
(Young modulus $E_{1}$ and $E_{2}$, and Poisson ratio $\nu _{1}$ and $\nu
_{2}$, respectively). If the root-mean-square slope of the roughness on both
surfaces is small one can show that this contact problem can be mapped on a
simpler problem with one elastic half space with a perfectly flat surface,
and another rigid solid with the combined roughness (see Fig. \ref{pich1.ps})%
\cite{Johnson}: 
\begin{equation*}
h(\mathbf{x})=h_{2}(\mathbf{x})-h_{1}(\mathbf{x}).
\end{equation*}%
The effective elastic module $E$ and Poisson ratio $\nu $ of the elastic
solid must satisfy 
\begin{equation*}
{\frac{1-\nu ^{2}}{E}}={\frac{1-\nu _{1}^{2}}{E_{1}}}+{\frac{1-\nu _{2}^{2}}{%
E_{2}}.}
\end{equation*}%
We will refer to this transformation as \textit{surface-mapping}. For
viscoelastic solids both the Young's modulus and the Poisson ratio depends
on the frequency $\omega$ (and hence, during sliding (velocity $v$), on the
length scale $v/\omega$). In this case the surface-mapping is no longer
valid, and in this paper we will show, within the Persson approach, how to
treat the more general problem of viscoelastic contact mechanics when both
solids have surface roughness.

\begin{figure}[tbp]
\begin{center}
\includegraphics[
                        width=0.49\textwidth,
                        angle=0]{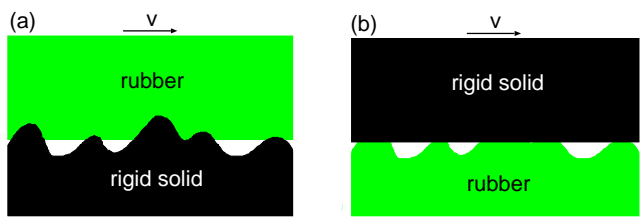}
\end{center}
\caption{(a) A rubber block with a smooth surface sliding relative to a
hard, randomly rough, substrate. (b) A rubber block with a rough surface
sliding relative to a hard solid with a flat surface. }
\label{twoblockpic.ps}
\end{figure}

\vskip 0.3cm \textbf{2 Qualitative discussion}

Let us consider a rubber block sliding on a rigid solid substrate. We assume
first that the rubber block has a perfectly flat surface and the rigid
substrate a randomly rough surface [see Fig. \ref{twoblockpic.ps}(a)]. In
this case during sliding the asperities of the rigid solid will induce
time-dependent deformations of the rubber, which results in the conversion
of translation energy into heat via the internal damping in the rubber. Thus
the viscoelastic deformations of the rubber will contribute to the kinetic
friction coefficient.

Let us now consider the opposite case where the rigid solid has a perfectly
smooth surface and the rubber block a rough surface [see Fig. \ref%
{twoblockpic.ps}(b)]. In this case during sliding the deformations of the
rubber does not change in time and no viscoelastic energy dissipation
occurs. Thus in this case there is no contribution to the friction from
viscoelastic deformations of the rubber. When roughness occurs on both
surfaces the situation is more complex but already the argument given here
indicates that the energy dissipation is dominated by the roughness in the
harder solid.

Another important difference between the two cases in Fig. \ref%
{twoblockpic.ps} relates to the area of real contact. In Fig. \ref%
{twoblockpic.ps}(a) the contact area depends on the viscoelastic modulus of
the rubber at finite frequencies. That is, during sliding at the speed $v $
a surface roughness component with wavelength $\lambda $ will result in
oscillating deformations of the rubber surface with a frequency $\omega
\approx v/\lambda $. Since the elastic modulus of rubber-like materials may
increase by a factor of $\sim 1000$ when the perturbing frequency increases
from the rubbery region to the glassy region, it is clear that a very strong
dependency of the contact area is expected: An increase in the elastic
modulus by a factor of $\sim 1000$ can result in a decrease in the contact
area by a factor of $\sim 1000$. However, when the surface roughness occurs
on the rubber surface while the hard counter-surface is perfectly flat,
there are no time-dependent viscoelastic deformations of the rubber and the
contact area is determined by the low-frequency viscoelastic modulus.

\begin{figure}[tbp]
\begin{center}
\includegraphics[
                        width=0.49\textwidth,
                        angle=0]{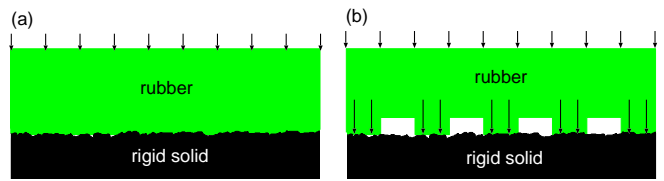}
\end{center}
\caption{If the contact area is small compared to the nominal contact area,
introducing a periodic roughness on the rubber block result in a negligible
change in the friction coefficient. Here we have assumed that adhesion does
not manifest itself macroscopically as a pull-off force. }
\label{blockfullpart.ps}
\end{figure}

Let us now compare the contact between a randomly rough hard solid (the
substrate) and (a) a rubber block with a flat surface [see Fig. \ref%
{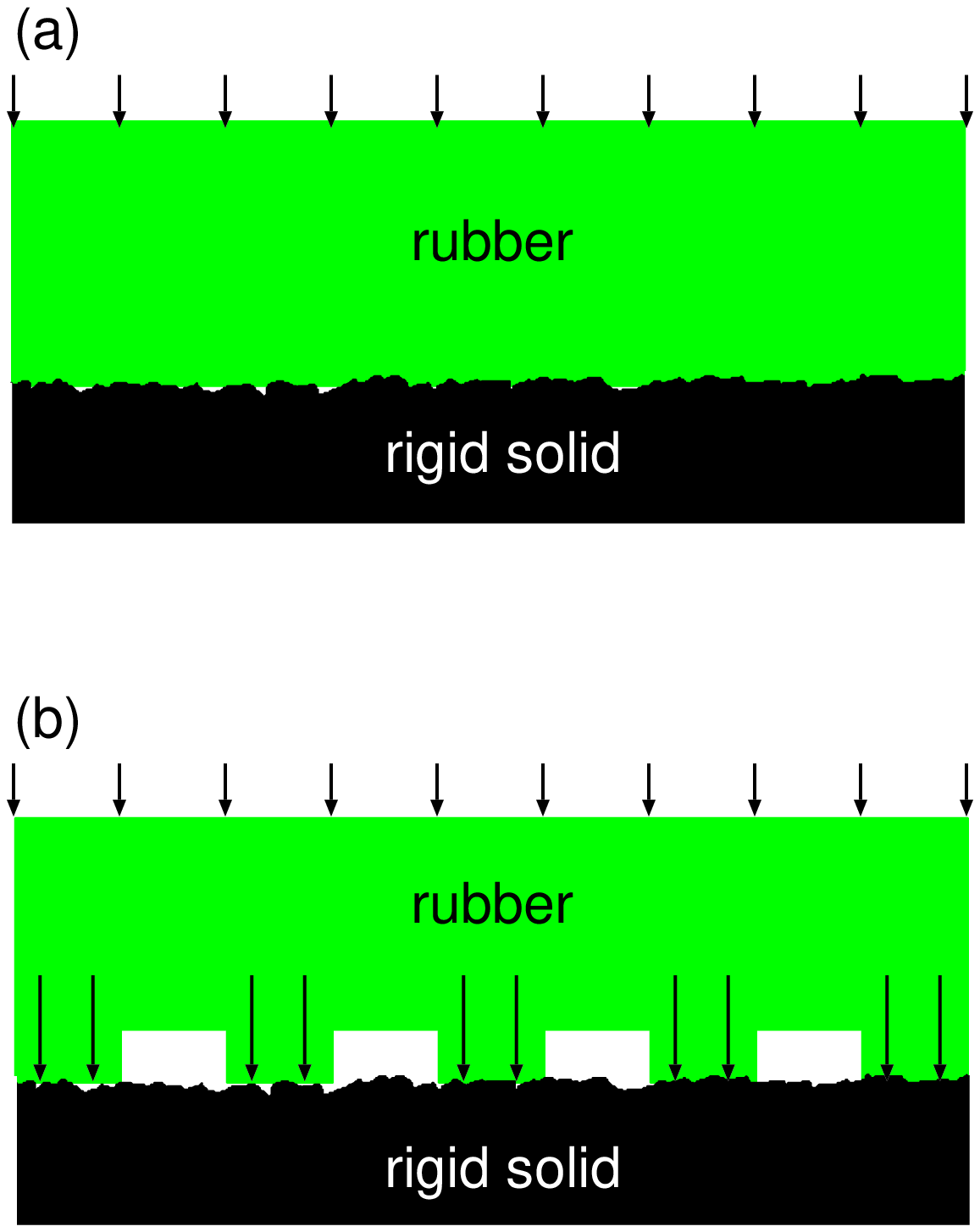}(a)], and (b) with a rubber block with a simple periodic
\textquotedblleft roughness\textquotedblright\ as indicated in Fig. \ref%
{blockfullpart.ps}(b). We assume that the longest wavelength roughness
component of the substrate surface is smaller than the linear size of the
(nominal) contact regions in \ref{blockfullpart.ps}(b). A uniform stress or
pressure $\sigma _{0}$ is assumed to act on the upper surface of the rubber
block. In case (a) a similar pressure is assumed to act at the interface
between the block and the substrate\cite{note}. According to the Persson
contact mechanics theory, if frictional heating can be neglected, the
viscoelastic contribution to the friction for small applied load is
proportional to the normal load, i.e. the friction coefficient is
independent of the load. Since the load is the same in both Fig. \ref%
{blockfullpart.ps}(a) and (b) it follows that the friction coefficient is
the same too. This is only true as long as adhesion does not manifest itself
macroscopically as a finite pull-off force, and as long as the contact area
is small compared to the nominal contact area. It is clear that if the
nominal contact area in case (b) is very small, the latter assumption will
no longer hold. But if all the assumptions made above holds, introducing a
periodic \textquotedblleft roughness\textquotedblright\ on the rubber
surface will have no influence on the friction coefficient and the contact
area. We will show in the next section that random roughness on the rubber
surface does influence the friction, but in most cases only to a very small
extent.

\vskip 0.3cm \textbf{3 Theory: special case}

In the following we will extend the contact mechanics and rubber friction
theory developed in Ref. \cite{Persson} to the case where both solids have
surface roughness, but one of those is rigid. In particular, in Sec. 3.1 we
consider the contact area and in Sec. 3.2 the rubber friction. The most
general case of arbitrary solids rheology is instead discussed in Sec. 5 and
in Appendix A, where we present analytical results for the contact area, the
rubber friction, and average interfacial separation.

\vskip 0.3cm \textbf{3.1 Contact area}

In this section we study the contact mechanics for the most important case
where solid \textbf{1} is viscoelastic (say rubber) and solid \textbf{2}
rigid, and both surfaces have surface roughness which are uncorrelated so
that $\langle h_{1}(\mathbf{x})h_{2}(\mathbf{x})\rangle =\langle h_{1}(%
\mathbf{x})\rangle \langle h_{2}(\mathbf{x})\rangle $ where $\langle
..\rangle $ stands for ensemble averaging. In this case, in order for solid 
\textbf{2} to make perfect contact with solid \textbf{1}, one needs to
deform the surface of solid \textbf{1} so that the normal surface
displacement $u_{\mathrm{z}}(\mathbf{x},t)$ equals the gap function:%
\begin{equation}
u_{\mathrm{z}}(\mathbf{x},t)=h_{2}(\mathbf{x},t)-h_{1}(\mathbf{x}).
\label{1}
\end{equation}%
Let us define the Fourier transform 
\begin{equation*}
u_{\mathrm{z}}(\mathbf{q},\omega )=(2\pi )^{-3}\int d^{2}xdt\ u_{\mathrm{z}}(%
\mathbf{x},t)e^{-i(\mathbf{q}\cdot \mathbf{x}-\omega t)}.
\end{equation*}%
If the upper solid moves at a constant velocity $\mathbf{v}=(v_{\mathrm{x}%
},v_{\mathrm{y}})$ parallel to the bottom solid, we have $h_{2}(\mathbf{x}%
,t)=h_{2}(\mathbf{x}-\mathbf{v}t)$ and 
\begin{equation*}
h_{2}(\mathbf{q},\omega )=h_{2}(\mathbf{q})\delta (\omega -\mathbf{q}\cdot 
\mathbf{v})
\end{equation*}%
and \ref{1} takes the form 
\begin{equation}
u_{\mathrm{z}}(\mathbf{q},\omega )=h_{2}(\mathbf{q})\delta (\omega -\mathbf{q%
}\cdot \mathbf{v})-h_{1}(\mathbf{q}).  \label{2}
\end{equation}%
If a normal stress or pressure $\sigma _{\mathrm{z}}(\mathbf{x},t)$ acts on
the surface of solid \textbf{1}, from the theory of viscoelasticity\cite%
{Persson} 
\begin{equation}
u_{\mathrm{z}}(\mathbf{q},\omega )=M(q,\omega )\sigma _{\mathrm{z}}(\mathbf{q%
},\omega ).  \label{3}
\end{equation}%
In the Persson contact mechanics theory the mean-square fluctuation of the
(normal) interfacial stress for complete contact (see Appendix A in Ref. 
\cite{Persson}) is needed to quantify the diffusion function:%
\begin{gather}
\langle \sigma ^{2}(\mathbf{x},t)\rangle ={\frac{(2\pi )^{3}}{t_{0}A_{0}}}%
\int d^{2}qd\omega \ |M(\mathbf{q},\omega )|^{-2}  \notag \\
\times \langle u_{\mathrm{z}}(\mathbf{q},\omega )u_{\mathrm{z}}(-\mathbf{q}%
,-\omega )\rangle ,  \label{4}
\end{gather}%
where $A_{0}$ is the surface area and $t_{0}$ the time over which we perform
averaging. Using \ref{2} we get%
\begin{gather*}
\left\langle u_{\mathrm{z}}(\mathbf{q},\omega )u_{\mathrm{z}}(-\mathbf{q}%
,-\omega )\right\rangle = \\
\langle h_{1}(\mathbf{q})h_{1}(-\mathbf{q})\rangle \lbrack \delta (\omega
)]^{2}+\langle h_{2}(\mathbf{q})h_{2}(-\mathbf{q})\rangle \lbrack \delta
(\omega -\mathbf{q}\cdot \mathbf{v})]^{2}
\end{gather*}%
where we have used that $h_{1}(\mathbf{q})$ and $h_{2}(\mathbf{q})$ are
(assumed to be) uncorrelated. Next, using that 
\begin{align}
\langle h_{1}(\mathbf{q})h_{1}(-\mathbf{q})\rangle &={\frac{A_{0}}{(2\pi
)^{2}}}C_{1}(\mathbf{q})  \label{5} \\
\langle h_{2}(\mathbf{q})h_{2}(-\mathbf{q})\rangle &={\frac{A_{0}}{(2\pi
)^{2}}}C_{2}(\mathbf{q})  \notag
\end{align}%
where $C_{1}(\mathbf{q})$ and $C_{2}(\mathbf{q})$ are the power spectra of
the surfaces of solids \textbf{1} and \textbf{2}, respectively, and that 
\begin{equation*}
\lbrack \delta (\omega )]^{2}=\delta (\omega ){\frac{t_{0}}{2\pi }}
\end{equation*}%
we get 
\begin{gather}
\langle u_{\mathrm{z}}(\mathbf{q},\omega )u_{\mathrm{z}}(-\mathbf{q},-\omega
)\rangle =  \notag \\
{\frac{A_{0}t_{0}}{(2\pi )^{3}}}\left[ C_{1}(\mathbf{q})\delta (\omega
)+C_{2}(\mathbf{q})\delta (\omega -\mathbf{q}\cdot \mathbf{v})\right] .
\label{6}
\end{gather}%
Substituting this result in \ref{4} gives 
\begin{gather}
\langle \sigma ^{2}(\mathbf{x},t)\rangle =\int d^{2}qd\omega \ |M(\mathbf{q}%
,\omega )|^{-2}  \notag \\
\times \left[ C_{1}(\mathbf{q})\delta (\omega )+C_{2}(\mathbf{q})\delta
(\omega -\mathbf{q}\cdot \mathbf{v})\right]  \notag \\
=\int d^{2}q\ \left[ |M(\mathbf{q},\mathbf{q}\cdot \mathbf{v})|^{-2}C_{2}(%
\mathbf{q})+|M(\mathbf{q},0)|^{-2}C_{1}(\mathbf{q})\right] .  \label{7}
\end{gather}%
For a homogeneous viscoelastic solid 
\begin{equation}
M^{-1}(\mathbf{q},\omega )=-\frac{1}{2}{\frac{E(\omega )}{1-\nu ^{2}\left(
\omega \right) }}q.  \label{8}
\end{equation}%
Substituting this in \ref{7} gives 
\begin{gather}
\langle \sigma ^{2}(\mathbf{x},t)\rangle ={\frac{1}{4}}\int d^{2}q\ q^{2} 
\notag \\
\times \left( \left\vert {\frac{E(\mathbf{q}\cdot \mathbf{v})}{1-\nu ^{2}(%
\mathbf{q}\cdot \mathbf{v})}}\right\vert ^{2}C_{2}(\mathbf{q})+\left\vert {%
\frac{E(0)}{1-\nu ^{2}(0)}}\right\vert ^{2}C_{1}(\mathbf{q})\right) .
\label{9}
\end{gather}%
If the rubber surface is smooth $C_{1}(\mathbf{q})=0$, \ref{9} reduces to
the expression derived in Ref. \cite{Persson}. For an elastic solid $%
E(\omega )=E(0)$ and $\nu (\omega )=\nu (0)$ are frequency independent so in
that case 
\begin{equation*}
\langle \sigma ^{2}(\mathbf{x},t)\rangle ={\frac{1}{4}}\int d^{2}q\ q^{2}%
\left[ {\frac{E}{1-\nu ^{2}}}\right] ^{2}[C_{2}(\mathbf{q})+C_{1}(\mathbf{q}%
)].
\end{equation*}%
This result is the expected one because for a rigid solid in contact with an
elastic solid the surface-mapping effective modulus $E/2(1-\nu ^{2})$ is
just that of the elastic solid, while the effective power spectrum is that
for the combined roughness given by 
\begin{equation*}
C(\mathbf{q})={\frac{(2\pi )^{2}}{A_{0}}}\langle \left[ h_{1}(\mathbf{q}%
)+h_{2}(\mathbf{q})\right] \left[ h_{1}(-\mathbf{q})+h_{2}(-\mathbf{q})%
\right] \rangle ,
\end{equation*}%
i.e. $C(\mathbf{q})=C_{1}(\mathbf{q})+C_{2}(\mathbf{q})$, where we again
have assumed uncorrelated roughness profiles. Thus, for elastic solids the
results above could be obtained directly using the surface-mapping.

In the Persson contact mechanics theory the relative contact area is given
by 
\begin{equation}
{\frac{A}{A_{0}}}=\mathrm{erf}\left( {\frac{1}{2\surd G}}\right) ,
\label{10}
\end{equation}%
where $\mathrm{erf}$ is the error function, with the integral representation 
\begin{equation*}
\mathrm{erf}(x)={\frac{2}{\surd \pi }}\int_{0}^{x}dx\ \mathrm{exp}(-x^{2}),
\end{equation*}%
and where 
\begin{gather}
G={\frac{1}{2\sigma _{0}^{2}}}\int d^{2}q\   \notag \\
\times \left[ |M(\mathbf{q},\mathbf{q}\cdot \mathbf{v})|^{-2}C_{2}(\mathbf{q}%
)+|M(\mathbf{q},0)|^{-2}C_{1}(\mathbf{q})\right] ,  \label{11}
\end{gather}%
where the integral is over all the roughness wavevectors. The nominal
contact pressure $\sigma _{0}=F_{0}/A_{0}$, where $F_{0}$ is the normal
force. For a homogeneous viscoelastic solid this expression takes the form 
\begin{equation*}
G={\frac{1}{8\sigma _{0}^{2}}}\int d^{2}q\ q^{2}\left[ \left\vert E_{\mathrm{%
r}}\left( \mathbf{q}\cdot \mathbf{v}\right) \right\vert ^{2}C_{2}(\mathbf{q}%
)+\left\vert {E_{\mathrm{r}}(0)}\right\vert ^{2}C_{1}(\mathbf{q})\right] ,
\end{equation*}%
where where $E_{\mathrm{r}}\left( \omega \right) =E\left( \omega \right) /%
\left[ 1-\nu ^{2}\left( \omega \right) \right] $. Note that $G$ depends on
the range of surface roughness included in the analysis. Thus, if $q_{0}$
correspond to the smallest surface roughness wavevector, and if we include
all the roughness $q_{0}<q<q_{1}$ then the function $G(q_{1})$ will mainly
depend on the cut-off wavevector $q_{1}$. In general, as $q_{1}$ increases, $%
G(q_{1})$ increases and the contact area $A(q_{1})$ decreases. For the
rubber friction to be discussed next we need the relative contact area $%
P(q)=A(q)/A_{0}$ for all $q$ between $q_{0}$ and $q_{1}$.

Using that for $x<<1$ we have $\mathrm{erf}(x)\approx 2x/\surd \pi $ we get
for large $G$ that $A/A_{0}\approx (\pi G)^{-1/2}$. Note that $G\rightarrow
\infty $ as $\sigma _{0}\rightarrow 0$ so for small nominal contact
pressures 
\begin{gather}
{\frac{A}{A_{0}}}=\frac{\sigma _{0}}{{E_{\mathrm{r}}(0)}}\left( {8/\pi }%
\right) ^{1/2}  \notag \\
\times \left\{ \int d^{2}q\ q^{2}\left[ \left( \frac{\left\vert E_{\mathrm{r}%
}\left( \mathbf{q}\cdot \mathbf{v}\right) \right\vert }{\left\vert {E_{%
\mathrm{r}}(0)}\right\vert }\right) ^{2}C_{2}(\mathbf{q})+C_{1}(\mathbf{q})%
\right] \right\} ^{-1/2}.  \label{12}
\end{gather}%
It is clear that the surface roughness on the solid 1 will reduce the
contact area.

\vskip 0.3cm \textbf{3.2 Rubber friction}

We now study the contribution to rubber friction from the viscoelastic
deformations of the rubber surface induced by the surface roughness of the
hard counter surface. Following Ref. \cite{Persson} we first focus on the
case of complete contact. In this case the dissipated energy is given by Eq. %
\ref{15} in Ref. \cite{Persson}: 
\begin{gather*}
\Delta E=\int d^{2}xdt\ \langle \dot{u}_{\mathrm{z}}(\mathbf{x},t)\sigma _{%
\mathrm{z}}(\mathbf{x},t)\rangle \\
=(2\pi )^{3}\int d^{2}qd\omega \ (i\omega )M^{-1}(\mathbf{q},\omega )\langle
u_{\mathrm{z}}(-\mathbf{q},-\omega )u_{\mathrm{z}}(\mathbf{q},\omega
)\rangle .
\end{gather*}%
Next using \ref{2} this equation takes the form 
\begin{gather}
\Delta E=A_{0}t_{0}\int d^{2}qd\omega \ (i\omega )M^{-1}(\mathbf{q},\omega )
\notag \\
\times \left[ C_{1}(\mathbf{q})\delta (\omega )+C_{2}(\mathbf{q})\delta
(\omega -\mathbf{q}\cdot \mathbf{v})\right]  \notag \\
=A_{0}t_{0}\int d^{2}q\ (i\mathbf{q}\cdot \mathbf{v})M^{-1}(\mathbf{q},%
\mathbf{q}\cdot \mathbf{v})C_{2}(\mathbf{q}).  \label{13}
\end{gather}%
As expected, for complete contact the dissipated energy does not depend on
the surface roughness on the rubber, but only on the roughness on the hard
counter-surface. As a result the rubber friction coefficient within the
Persson approach is given by the same expression as derived in Refs. \cite%
{Persson,c9}, except now enters the contact area function $P(q)$ given by %
\ref{10} and \ref{11}, rather than the expression derived in Ref. \cite%
{Persson} for a perfectly smooth rubber surface: 
\begin{gather}
\mu ={\frac{1}{2}}\int_{q_{0}}^{q_{1}}dq\ q^{3}C_{2}(q)P(q)S(q)  \notag \\
\times \int_{0}^{2\pi }d\phi \ \mathrm{cos}\phi \ \mathrm{Im}{\frac{E(\omega
)}{\sigma _{0}(1-\nu ^{2})}}  \label{14}
\end{gather}%
where $P(q)=A(q)/A_{0}$ is given by \ref{10} and where the correction factor 
\cite{c9}%
\begin{equation*}
S(q)=\gamma +(1-\gamma )P^{2}(q)
\end{equation*}%
with $\gamma \approx 1/2$. Note that $P(q)$ [and $S(q)$] depend on the
applied (or nominal) pressure $\sigma _{0}$ and sometimes, when necessary,
we write $P(q)=P(q,\sigma _{0})$.

\begin{figure}[tbp]
\begin{center}
\includegraphics[
                        width=0.45\textwidth,
                        angle=0]{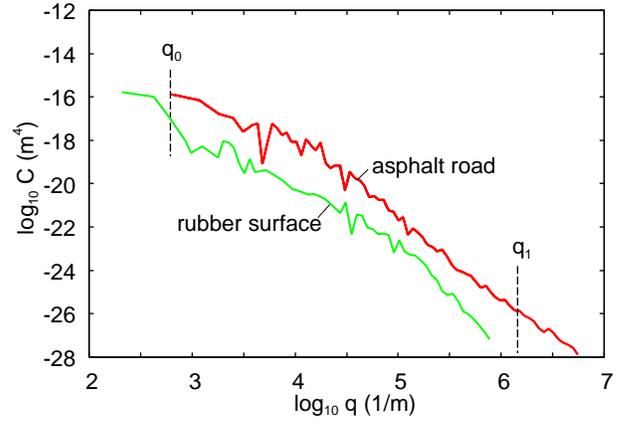}
\end{center}
\caption{The surface roughness power spectra of an asphalt road surface and
a rubber tread block surface after use. }
\label{1logq.2logC.ps}
\end{figure}

\begin{figure}[tbp]
\begin{center}
\includegraphics[
                        width=0.45\textwidth,
                        angle=0]{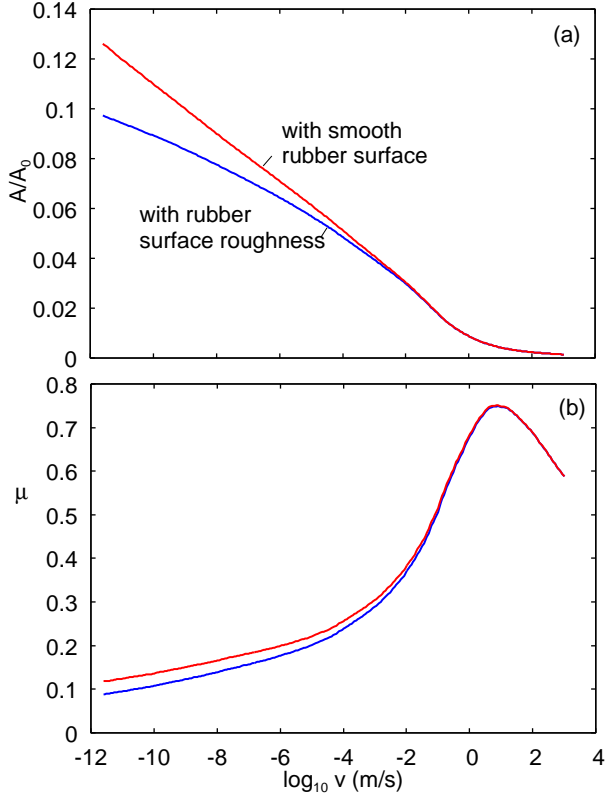}
\end{center}
\caption{The calculated contact area (a) and friction coefficient (b) for a
rubber tread block sliding on a road surface with the surface roughness
power spectrum shown in Fig. \protect\ref{1logq.2logC.ps} (red curve). The
red curve is assuming the rubber surface is completely flat (no surface
roughness) and the blue curve assuming the same surface roughness on the
rubber surface as on the road surface. If one instead use the actual surface
roughness on the tread block (green line in Fig. \protect\ref{1logq.2logC.ps}%
) the calculated friction coefficient and contact area becomes so similar to
the case of the flat rubber surface that it overlap the red curves. For the
nominal contact pressure $\protect\sigma _{0}=0.15\ \mathrm{MPa}$ and the
temperature $T=20^{\circ }\mathrm{C}$. Calculations are without including
the effect of frictional heating.}
\label{1logv.2mu.with.without.rubber.roughness.ps}
\end{figure}

\vskip 0.3cm \textbf{4 Numerical results}

In this section we will illustrate the theory presented above with an
example. Fig. \ref{1logq.2logC.ps} shows the surface roughness power spectra
of an asphalt road surface and a rubber tread block surface after use. Note
that, as expected, the road surface exhibits much larger roughness then the
tread block.

By using the theory above one finds that the friction (and the contact area)
between the asphalt road surface and the rubber surface with the power
spectra given in Fig. \ref{1logq.2logC.ps} are nearly identical to the case
when the rubber surface is perfectly flat. For this reason, we instead
assume that the rubber surface is as rough as the asphalt road surface i.e.
it has a power spectrum given by the red curve in Fig. \ref{1logq.2logC.ps}.
For this case we show in Fig. \ref%
{1logv.2mu.with.without.rubber.roughness.ps} (blue curves) the calculated
contact area (a) and friction coefficient (b). The red curves in the same
figure are the results when the rubber surface is completely flat (no
surface roughness). It is remarkable how small influence the surface
roughness on the rubber block would have on the contact area and the
friction even in this extreme case. Note in particular that for high sliding
speeds the roughness on the rubber surface has negligible influence on the
calculated friction. This is easy to understand from \ref{12} and \ref{13}.
Indeed, the contact area depends on the sum 
\begin{equation}
\int d^{2}q\ q^{2}\left( \left\vert {\frac{E_{\mathrm{r}}(\mathbf{q}\cdot 
\mathbf{v})}{\sigma _{0}}}\right\vert ^{2}C_{2}(\mathbf{q})+\left\vert {%
\frac{E_{\mathrm{r}}(0)}{\sigma _{0}}}\right\vert ^{2}C_{1}(\mathbf{q}%
)\right) .  \label{15}
\end{equation}%
For small sliding speeds, $\omega =\mathbf{q}\cdot \mathbf{v}$ is very small
in a large part of the \textbf{q}-plane, and the $\omega =\mathbf{q}\cdot 
\mathbf{v}$ and $\omega =0$ terms in \ref{15} are of nearly the same
magnitude in a large part of the $\mathbf{q}$-integration domain. However,
for high sliding speeds $\omega =\mathbf{q}\cdot \mathbf{v}$ will be very
large in a large part of the $\mathbf{q}$-plane, and since $E(\omega )$
increases strongly as the frequency increases from the rubbery region (small 
$\omega $) to the glassy region (large $\omega $), it is clear that for
large sliding speeds the $\omega =\mathbf{q}\cdot \mathbf{v}$ term in \ref%
{15} will be much larger than the $\omega =0$ term in a large region of the $%
\mathbf{q}$-plane. This explain why the two curves in Fig. \ref%
{1logv.2mu.with.without.rubber.roughness.ps} (a) and (b) approach each other
as the velocity $v$ increases.

\vskip0.3cm \textbf{5 Theory: general case}

We now consider the general case where both solids have surface roughness
and arbitrary viscoelastic properties. Using the Persson contact mechanics
theory, in Appendix A we derive expressions for the contact area, the
viscoelastic contribution to the sliding friction, and the average
interfacial separation. In this section we consider 4 limiting cases (see
Fig. \ref{All4Cases.ps}) of the full theory.

We first summarize the basic results obtained in Appendix A. We assume that
solid \textbf{1} moves with the velocity $\mathbf{v}_{1}$ and solid \textbf{2%
} with the velocity $\mathbf{v}_{2}$. All physical properties depends only
on the velocity difference $\mathbf{v}=\mathbf{v}_{2}-\mathbf{v}_{1}$ so we
can always assume $\mathbf{v}_{2}=\mathbf{v}$ and $\mathrm{v}_{1}=\mathbf{0}$%
. The contact area is given by \ref{10} with: 
\begin{equation}
G\left( q_{1}\right) ={\frac{1}{2\sigma _{0}^{2}}}\int_{q_{0}}^{q_{1}}d^{2}q%
\sum_{j=1}^{2}{\ \frac{C_{j}\left( \mathbf{q}\right) }{|M\left( \mathbf{q},%
\mathbf{q}\cdot \mathbf{v}_{j}\right) |^{2}}},  \label{xA11}
\end{equation}%
where 
\begin{equation}
M\left( \mathbf{q},\omega \right) =M_{1}\left( \mathbf{q},\omega -\mathbf{q}%
\cdot \mathbf{v}_{1}\right) +M_{2}\left( \mathbf{q},\omega -\mathbf{q}\cdot 
\mathbf{v}_{2}\right) .  \label{xA6}
\end{equation}%
The viscoelastic contribution to the friction coefficient is given by: 
%\begin{equation}
\begin{equation}
\mu =\int d^{2}q~S\left( q\right) P\left( q\right) \sum_{j=1}^{2}\frac{\dot{M%
}\left( \mathbf{q},\mathbf{q}\cdot \mathbf{v}_{j}\right) C_{j}\left( \mathbf{%
q}\right) }{\sigma _{0}|M\left( \mathbf{q},\mathbf{q}\cdot \mathbf{v}%
_{j}\right) |^{2}},  \label{xA17}
\end{equation}%
where 
\begin{equation*}
\dot{M}\left( \mathbf{q},\omega \right) =\dot{M}_{1}\left( \mathbf{q},\omega
-\mathbf{q}\cdot \mathbf{v}_{1}\right) +\dot{M}_{2}\left( \mathbf{q},\omega -%
\mathbf{q}\cdot \mathbf{v}_{2}\right)
\end{equation*}%
and%
\begin{equation*}
\dot{M}_{j}\left( \mathbf{q},\omega \right) =i\omega M_{j}\left( \mathbf{q}%
,\omega \right) .
\end{equation*}%
The average interfacial separation is given by: %\begin{equation}
\begin{gather}
\bar{u}\left( \sigma _{0}\right) =\int d^{2}q\sum_{j=1}^{2}{C_{j}\left( 
\mathbf{q}\right) \mathrm{Re}\left[ M\left( \mathbf{q},\mathbf{q}\cdot \mathbf{%
v}_{j}\right) ^{-1}\right] }  \notag \\
\times \int_{\sigma _{0}}^{\infty }dp\frac{1}{2p}\frac{\partial \left[
S\left( q,p\right) P\left( q,p\right) \right] }{\partial p}%
.  \label{xA18}
\end{gather}

\begin{figure}[tbp]
\begin{center}
\includegraphics[
                        width=0.5\textwidth,
                        angle=0]{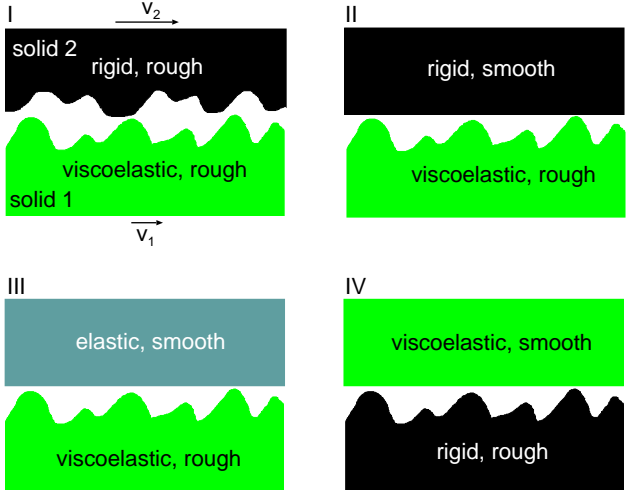}
\end{center}
\caption{Four limiting cases discussed in the text.}
\label{All4Cases.ps}
\end{figure}

\vskip 0.1cm \textbf{Case I}: \textit{A rigid solid \textbf{2} sliding on a
viscoelastic solid \textbf{1}. Both solids have random surface roughness.}

This limiting case is the same as the theory developed in Sec. 3. With $%
M_{2}=0$, and considering that $\mathbf{v}=\mathbf{v}_{2}-\mathbf{v}_{1}$,
we have%
\begin{align*}
M\left( \mathbf{q},\omega \right) &=M_{1}\left( \mathbf{q},\omega -\mathbf{q}%
\cdot \mathbf{v}_{1}\right) \\
\dot{M}\left( \mathbf{q},\omega \right) &=i\left( \omega -\mathbf{q}\cdot 
\mathbf{v}_{1}\right) M\left( \mathbf{q},\omega \right) .
\end{align*}

The contact area depends on [see \ref{xA11}]: 
\begin{equation*}
\sum_{j=1}^{2}\frac{C_{j}\left( \mathbf{q}\right) }{\left\vert M\left( 
\mathbf{q},\mathbf{q}\cdot \mathbf{v}_{j}\right) \right\vert ^{2}}={\frac{%
C_{1}\left( \mathbf{q}\right) }{\left\vert M_{1}\left( \mathbf{q},0\right)
\right\vert ^{2}}+\frac{C_{2}\left( \mathbf{q}\right) }{\left\vert
M_{1}\left( \mathbf{q},\mathbf{q}\cdot \mathbf{v}\right) \right\vert ^{2}}.}
\end{equation*}%
Note that this equation correspond to \ref{11} as expected.

Similarly the viscoelastic contribution to the friction depends on [see \ref%
{xA17}]: 
\begin{gather}
\sum_{j=1}^{2}{\frac{\dot{M}\left( \mathbf{q},\mathbf{q}\cdot \mathbf{v}%
_{j}\right) C_{j}\left( \mathbf{q}\right) }{\left\vert M\left( \mathbf{q},%
\mathbf{q}\cdot \mathbf{v}_{j}\right) \right\vert ^{2}}}  \notag \\
=\sum_{j=1}^{2}{\frac{i\left( \mathbf{q}\cdot \mathbf{v}_{j}-\mathbf{q}\cdot 
\mathbf{v}_{1}\right) C_{j}\left( \mathbf{q}\right) }{M_{1}\left( -\mathbf{q}%
,\mathbf{-q}\cdot \left( \mathbf{v}_{j}-\mathbf{v}_{1}\right) \right) }=%
\frac{i\mathbf{q}\cdot \mathbf{v}C_{2}\left( \mathbf{q}\right) }{M_{1}\left(
-\mathbf{q},\mathbf{-q}\cdot \mathbf{v}\right) }.}  \label{17}
\end{gather}%
Note that \ref{17} correspond to \ref{13} as expected.

Finally, the average interfacial separation depends on [see \ref{xA18}]: 
\begin{equation}
\sum_{j=1}^{2}{\frac{C_{j}\left( \mathbf{q}\right) }{M\left( \mathbf{q},%
\mathbf{q}\cdot \mathbf{v}_{j}\right) }=\frac{C_{1}\left( \mathbf{q}\right) 
}{M_{1}\left( \mathbf{q},0\right) }+\frac{C_{2}\left( \mathbf{q}\right) }{%
M_{1}\left( \mathbf{q},\mathbf{q}\cdot \mathbf{v}\right) }.}  \label{18}
\end{equation}

As discussed in Sec. 4, we note that, at least for high sliding speed, the
power spectrum of the compliant solid usually plays a negligible role in the
generation of the contact area. Similar, the average interface separation $%
\bar{u}$ is determined mainly by the power spectrum of the rigid solid, but
note the difference in the potency of the $M_1$-term (two for the contact
area and one for the average separation).

\vskip 0.1cm \textbf{Case II}: \textit{A rigid solid \textbf{2} with smooth
surface sliding on a viscoelastic solid \textbf{1} with rough surface.}

With $M_{2}=0$ and $h_{2}=0$, we have%
\begin{align*}
M\left( \mathbf{q},\omega \right) &=M_{1}\left( \mathbf{q},\omega -\mathbf{q}%
\cdot \mathbf{v}_{1}\right) \\
\dot{M}\left( \mathbf{q},\omega \right) &=i\left( \omega -\mathbf{q}\cdot 
\mathbf{v}_{1}\right) M\left( \mathbf{q},\omega \right) .
\end{align*}

The contact area depends on: 
\begin{equation}
\sum_{j=1}^{2}\frac{C_{j}\left( \mathbf{q}\right) }{\left\vert M\left( 
\mathbf{q},\mathbf{q}\cdot \mathbf{v}_{j}\right) \right\vert ^{2}}{=\frac{%
C_{1}\left( \mathbf{q}\right) }{\left\vert M_{1}\left( \mathbf{q},0\right)
\right\vert ^{2}}.}  \label{19}
\end{equation}

The viscoelastic contribution to the friction depends on: 
\begin{equation}
\sum_{j=1}^{2}{\frac{\dot{M}\left( \mathbf{q},\mathbf{q}\cdot \mathbf{v}%
_{j}\right) C_{j}\left( \mathbf{q}\right) }{\left\vert M\left( \mathbf{q},%
\mathbf{q}\cdot \mathbf{v}_{j}\right) \right\vert ^{2}}=}0.  \label{20}
\end{equation}%
Hence, as expected, no hysteresis friction is generated during the sliding
interaction. This is in accordance with the qualitative discussion presented
in Sec. 2.

The average interfacial separation depends on: 
\begin{equation}
\sum_{j=1}^{2}{\frac{C_{j}\left( \mathbf{q}\right) }{M\left( \mathbf{q},%
\mathbf{q}\cdot \mathbf{v}_{j}\right) }=\frac{C_{1}\left( \mathbf{q}\right) 
}{M_{1}\left( \mathbf{q},0\right) }.}  \label{21}
\end{equation}%
Note that both the contact area and the average interfacial separation
depends on the viscoelastic modulus at vanishing frequency (i.e., in the
rubbery regime). This is of course expected as there are no time-dependent
deformations of the rubber surface.

\vskip 0.1cm \textbf{Case III}: \textit{An elastic solid \textbf{2} with a
smooth surface sliding on a viscoelastic solid \textbf{1} with rough surface.%
}

With $M_{2}\left( \mathbf{q},\omega \right) =M_{2}\left( \mathbf{q}\right) $
and $h_{2}=0$, we have%
\begin{align*}
\dot{M}\left( \mathbf{q},\omega \right) &=\dot{M}_{1}\left( \mathbf{q}%
,\omega -\mathbf{q}\cdot \mathbf{v}_{1}\right) +\dot{M}_{2}\left( \mathbf{q}%
,\omega -\mathbf{q}\cdot \mathbf{v}_{2}\right) \\
M\left( \mathbf{q},\omega \right) &=M_{1}\left( \mathbf{q},\omega -\mathbf{q}%
\cdot \mathbf{v}_{1}\right) +M_{2}\left( \mathbf{q}\right) .
\end{align*}

The contact area depends on: 
\begin{equation}
\sum_{j=1}^{2}\frac{C_{j}\left( \mathbf{q}\right) }{\left\vert M\left( 
\mathbf{q},\mathbf{q}\cdot \mathbf{v}_{j}\right) \right\vert ^{2}}{=\frac{%
C_{1}\left( \mathbf{q}\right) }{\left\vert M_{1}\left( \mathbf{q},0\right)
+M_{2}\left( \mathbf{q}\right) \right\vert ^{2}}.}  \label{22}
\end{equation}

The viscoelastic contribution to the friction depends on: 
\begin{equation}
\sum_{j=1}^{2}{\frac{\dot{M}\left( \mathbf{q},\mathbf{q}\cdot \mathbf{v}%
_{j}\right) C_{j}\left( \mathbf{q}\right) }{\left\vert M\left( \mathbf{q},%
\mathbf{q}\cdot \mathbf{v}_{j}\right) \right\vert ^{2}}=\frac{-i\left( 
\mathbf{q}\cdot \mathbf{v}\right) M_{2}\left( \mathbf{q}\right) C_{1}\left( 
\mathbf{q}\right) }{\left\vert M_{1}\left( \mathbf{q},0\right) +M_{2}\left( 
\mathbf{q}\right) \right\vert ^{2}}.}  \label{23}
\end{equation}

The average interfacial separation depends on: 
\begin{equation}
\sum_{j=1}^{2}{\frac{C_{j}\left( \mathbf{q}\right) }{M\left( \mathbf{q},%
\mathbf{q}\cdot \mathbf{v}_{j}\right) }=\frac{C_{1}\left( \mathbf{q}\right) 
}{M_{1}\left( \mathbf{q},0\right) +M_{2}\left( \mathbf{q}\right) }.}
\label{24}
\end{equation}%
We note first in \ref{22} and \ref{24} that the equivalent interfacial
compliance, related to the contact area as well as to the average
interfacial separation, is determined by a simple summation rule of each
solid compliance, evaluated under static interaction. Moreover, even if \ref%
{23} is non-zero, $\left( \mathbf{q}\cdot \mathbf{v}\right) M_{2}\left( 
\mathbf{q}\right) $ is an odd function of $\mathbf{q}$, resulting into a
zero micro rolling friction [see \ref{A17}].

\vskip 0.1cm \textbf{Case IV}: \textit{A viscoelastic solid \textbf{2} with
a smooth surface sliding on a rigid solid \textbf{1} with rough surface.}

With $M_{1}=0$ and $h_{2}=0$ we have%
\begin{align*}
M\left( \mathbf{q},\omega \right) &=M_{2}\left( \mathbf{q},\omega -\mathbf{q}%
\cdot \mathbf{v}_{2}\right) \\
\dot{M}\left( \mathbf{q},\omega \right) &=i\left( \omega -\mathbf{q}\cdot 
\mathbf{v}_{2}\right) M\left( \mathbf{q},\omega \right) .
\end{align*}

The contact area depends on: 
\begin{equation}
\sum_{j=1}^{2}\frac{C_{j}\left( \mathbf{q}\right) }{\left\vert M\left( 
\mathbf{q},\mathbf{q}\cdot \mathbf{v}_{j}\right) \right\vert ^{2}}{=\frac{%
C_{1}\left( \mathbf{q}\right) }{\left\vert M_{2}\left( \mathbf{q},-\mathbf{q}%
\cdot \mathbf{v}\right) \right\vert ^{2}}.}  \label{25}
\end{equation}

The viscoelastic contribution to the friction depends on: 
\begin{equation}
\sum_{j=1}^{2}{\frac{\dot{M}\left( \mathbf{q},\mathbf{q}\cdot \mathbf{v}%
_{j}\right) C_{j}\left( \mathbf{q}\right) }{\left\vert M\left( \mathbf{q},%
\mathbf{q}\cdot \mathbf{v}_{j}\right) \right\vert ^{2}}=-\frac{i\left( 
\mathbf{q}\cdot \mathbf{v}\right) C_{1}\left( \mathbf{q}\right) }{%
M_{2}\left( -\mathbf{q},\mathbf{q}\cdot \mathbf{v}\right) }.}  \label{26}
\end{equation}%
Assuming $\mathbf{v}=\mathbf{v}_{2}-\mathbf{v}_{1}$ is aligned along the 
\textrm{x}-axis, using \ref{26} in \ref{xA17} we obtain%
\begin{equation*}
\mu =\int d^{2}q~S\left( q\right) P\left( q\right) \frac{iq_{x}}{\sigma
_{0}M_{2}\left( -\mathbf{q},q_{x}v\right) }C_{1}\left( \mathbf{q}\right) ,
\end{equation*}%
which for a viscoelastic half space reads (removing subscripts for
simplicity) \cite{c9}%
\begin{equation*}
\mu \left( \sigma _{0}\right) =\frac{1}{2\sigma _{0}}\int
d^{2}q~q_{x}qS\left( q\right) P\left( q\right) C_{1}\left( \mathbf{q}\right) 
\mathrm{Im}\left[ E_{\mathrm{r}}\left( q_{x}v\right) \right]
\end{equation*}%
which corresponds to \ref{14}.

The average interfacial separation depends on: 
\begin{equation}
\sum_{j=1}^{2}{\frac{C_{j}\left( \mathbf{q}\right) }{M\left( \mathbf{q},%
\mathbf{q}\cdot \mathbf{v}_{j}\right) }=\frac{C_{1}\left( \mathbf{q}\right) 
}{M_{2}\left( \mathbf{q},-\mathbf{q}\cdot \mathbf{v}\right) }.}  \label{27}
\end{equation}

\vskip 0.3cm \textbf{6 Summary and conclusion}

We have studied the contact mechanics and sliding friction between a
viscoelastic solid and a rigid solid where both objects have random
uncorrelated roughness. We have shown that for a rubber-like material, where
the Young's modulus may increase by a factor $\sim 1000$ when going from the
rubbery region to the glassy region, roughness on the rubber surface will in
most cases have a rather small influence on the sliding (velocity $v$)
contact. However, for static contact, obtained in the limit $v\rightarrow 0$%
, the roughness on both surfaces is equally important. As a numerical
application we considered the contact between a rubber tread block and a
road surface, and showed that during sliding the surface roughness on the
rubber surface has in most cases only a very small influence on the contact
area and the viscoelastic sliding friction coefficient.

We have also studied the general case where both solids have random
roughness and arbitrary (linear) rheology properties. For this case, within
the Persson contact mechanics approach, we have derived expressions for the
contact area, the viscoelastic contribution to the friction force, and for
the average surface separation. We have studied 4 different limiting cases
of the theory.

The theory presented in this paper can be extended to correlated surface
roughness, and to include frictional heating and adhesion.

\onecolumngrid
\appendix
\section{General contact mechanics theory}
Here we derive the general contact mechanics theory for steady sliding rough
solids (to which a moving reference $\mathbf{\xi }_{i}$ is associated, where 
$i=1$ for the lower solid, see Fig. \ref{pich1.ps}), in adhesive- and
friction-less interaction. We define 
\begin{equation}
u_{\mathrm{z}}\left( \mathbf{q},z,\omega \right) =\left( 2\pi \right)
^{-3}\int dt\int d^{2}x~u_{\mathrm{z}}\left( \mathbf{x},z,t\right)
e^{-i\left( \mathbf{q}\cdot \mathbf{x}-\omega t\right) },  \label{A00}
\end{equation}%
where $u_{\mathrm{z}}\left( \mathbf{x},z,t\right) $ is the generic surface
out-of-plane displacement, and 
\begin{equation}
\sigma \left( \mathbf{q},z,\omega \right) =\left( 2\pi \right) ^{-3}\int
dt\int d^{2}x~\sigma \left( \mathbf{x},z,t\right) e^{-i\left( \mathbf{q}%
\cdot \mathbf{x}-\omega t\right) },  \label{A01}
\end{equation}%
where $\sigma \left( \mathbf{x},z=0,t\right) =\sigma \left( \mathbf{x}%
,t\right) $ is the contact stress, both observed in the fixed reference $%
\mathbf{\xi }$. Hence, given that $\mathbf{\xi }_{j}=\mathbf{\xi }-\mathbf{v}%
_{j}t$ [resulting into $\sigma _{|\xi }\left( \mathbf{q},\omega \right)
=\sigma _{|\xi _{j}}\left( \mathbf{q},\omega -\mathbf{q}\cdot \mathbf{v}%
_{j}\right) $, and similarly for $u_{\mathrm{z}}$], we get%
\begin{equation}
\sigma \left( \mathbf{q},\omega \right) =M_{j}^{-1}\left( \mathbf{q},\omega -%
\mathbf{q}\cdot \mathbf{v}_{j}\right) u_{\mathrm{z}j}\left( \mathbf{q}%
,\omega \right) ,  \label{A1}
\end{equation}%
whereas in term of the displacement time derivatives [note: $\frac{d}{%
dt_{|\xi _{j}}}\left[ u_{\mathrm{z}j}\left( \mathbf{q},\omega \right) \right]
=\frac{d}{dt_{|\xi }}\left[ u_{\mathrm{z}j}\left( \mathbf{q},\omega \right) %
\right] +u_{\mathrm{z}j}\left( \mathbf{q},\omega \right) i\mathbf{q}\cdot 
\mathbf{v}_{j}$, resulting into $i\left( \omega -\mathbf{q}\cdot \mathbf{v}%
_{j}\right) u_{\mathrm{z}j}\left( \mathbf{q},\omega \right) =\dot{u}_{%
\mathrm{z}j}\left( \mathbf{q},\omega \right) $]%
\begin{equation}
\sigma \left( \mathbf{q},\omega \right) =\dot{M}_{j}^{-1}\left( \mathbf{q}%
,\omega -\mathbf{q}\cdot \mathbf{v}_{j}\right) \dot{w}_{j}\left( \mathbf{q}%
,\omega \right) ,  \label{A2}
\end{equation}%
where simply%
\begin{equation}
\dot{M}_{j}\left( \mathbf{q},\omega \right) =i\omega M_{j}\left( \mathbf{q}%
,\omega \right) .  \label{A3}
\end{equation}%
Hence, by equating the stress in \ref{A1} for both top and bottom solid we
get%
\begin{equation}
u_{\mathrm{z}2}\left( \mathbf{q},\omega \right) =M_{2}\left( \mathbf{q}%
,\omega -\mathbf{q}\cdot \mathbf{v}_{2}\right) M_{1}^{-1}\left( \mathbf{q}%
,\omega -\mathbf{q}\cdot \mathbf{v}_{1}\right) u_{\mathrm{z}1}\left( \mathbf{%
q},\omega \right)   \label{A4}
\end{equation}%
so that%
\begin{equation}
u_{\mathrm{z}}\left( \mathbf{q},\omega \right) =u_{\mathrm{z}1}\left( 
\mathbf{q},\omega \right) +u_{\mathrm{z}2}\left( \mathbf{q},\omega \right) =%
\left[ 1+M_{2}\left( \mathbf{q},\omega -\mathbf{q}\cdot \mathbf{v}%
_{2}\right) M_{1}^{-1}\left( \mathbf{q},\omega -\mathbf{q}\cdot \mathbf{v}%
_{1}\right) \right] u_{\mathrm{z}1}\left( \mathbf{q},\omega \right) .
\label{A5}
\end{equation}%
We can determine the equivalent interface compliance\ $M\left( \mathbf{q}%
,\omega \right) $ from%
\begin{equation*}
\sigma \left( \mathbf{q},\omega \right) =M^{-1}\left( \mathbf{q},\omega \right)
u_{\mathrm{z}}\left( \mathbf{q},\omega \right) ,
\end{equation*}%
where%
\begin{equation}
M\left( \mathbf{q},\omega \right) =M_{1}\left( \mathbf{q},\omega -\mathbf{q}%
\cdot \mathbf{v}_{1}\right) +M_{2}\left( \mathbf{q},\omega -\mathbf{q}\cdot 
\mathbf{v}_{2}\right) .  \label{A6}
\end{equation}%
Moreover since $\dot{u}_{\mathrm{z}i}\left( \mathbf{q},\omega \right) /\dot{M%
}_{i}\left( \mathbf{q},\omega -\mathbf{q}\cdot \mathbf{v}_{i}\right) =u_{%
\mathrm{z}}\left( \mathbf{q},\omega \right) /M\left( \mathbf{q},\omega
\right) $ we can determine the equivalent interface compliance\ derivative $%
\dot{M}\left( \mathbf{q},\omega \right) $ from%
\begin{equation*}
\sigma \left( \mathbf{q},\omega \right) =\dot{M}^{-1}\left( \mathbf{q},\omega
\right)\dot{u}_{\mathrm{z}}\left( \mathbf{q},\omega \right) ,
\end{equation*}%
where%
\begin{equation}
\dot{M}\left( \mathbf{q},\omega \right) =\dot{M}_{1}\left( \mathbf{q},\omega
-\mathbf{q}\cdot \mathbf{v}_{1}\right) +\dot{M}_{2}\left( \mathbf{q},\omega -%
\mathbf{q}\cdot \mathbf{v}_{2}\right) .  \label{A7}
\end{equation}%
Observe finally that from \ref{A00} and \ref{A01} it follows 
\begin{align*}
\dot{M}^{-1}\left( \mathbf{q},\omega \right)\dot{M}^{-1}\left( -\mathbf{q}%
,-\omega \right) & =\left\vert \dot{M}\left( \mathbf{q},\omega \right)
\right\vert ^{-2} \\
M^{-1}\left( \mathbf{q},\omega \right) M^{-1}\left( -\mathbf{q},-\omega \right)
& =\left\vert M\left( \mathbf{q},\omega \right)\right\vert ^{-2}
\end{align*}%
i.e. $\left[ M^{-1}\left( -\mathbf{q},-\omega \right)\right] =\left[
M^{-1}\left( \mathbf{q},\omega \right) \right] ^{\ast }$ [and similalry for $%
\dot{M}^{-1}\left( \mathbf{q},\omega \right)$].

We calculate first the stress auto-correlation for full contact. In such a
case%
\begin{equation*}
u_{\mathrm{z}}\left( \mathbf{q},\omega \right) =\sum_{j=1}^{2}{h_{j}\left( 
\mathbf{q}\right) \delta \left( \omega -\mathbf{q}\cdot \mathbf{v}%
_{j}\right) }
\end{equation*}%
i.e.%
\begin{equation*}
\sigma \left( \mathbf{q},\omega \right) =M^{-1}\left( \mathbf{q},\omega \right)
\sum_{j=1}^{2}{h_{j}\left( \mathbf{q}\right) \delta \left( \omega -%
\mathbf{q}\cdot \mathbf{v}_{j}\right) }.
\end{equation*}%
Hence, considering that $C_{j}\left( \mathbf{q}\right) \delta \left( \mathbf{%
q}+\mathbf{\bar{q}}\right) =\left\langle h_{j}\left( \mathbf{q}\right)
h_{j}\left( \mathbf{\bar{q}}\right) \right\rangle $ and assuming the surface
roughness are uncorrelated%
\begin{align}
\left\langle \sigma ^{2}\left( \mathbf{x},t\right) \right\rangle &=\int
d^{2}qd\omega \int d^{2}\bar{q}d\bar{\omega}~\left\langle \sigma \left( 
\mathbf{q},\omega \right) \sigma \left( \mathbf{\bar{q}},\bar{\omega}\right)
\right\rangle e^{i\left( \mathbf{q}\cdot \mathbf{x}-\omega t\right)
}e^{i\left( \mathbf{\bar{q}}\cdot \mathbf{x}-\bar{\omega}t\right) }  \notag
\\
&=\int d^{2}q\sum_{j=1}^{2}{\frac{C_{j}\left( \mathbf{q}\right) }{\left\vert
M\left( \mathbf{q},\mathbf{q}\cdot \mathbf{v}_{j}\right) \right\vert ^{2}}}.
\label{A8}
\end{align}%
With similar considerations we can calculate the correlation function $%
\left\langle \sigma \left( \mathbf{x},t\right) \dot{u}_{\mathrm{z}}\left( 
\mathbf{x},t\right) \right\rangle $%
\begin{align}
\left\langle \sigma \left( \mathbf{x},t\right) \dot{u}_{\mathrm{z}}\left( 
\mathbf{x},t\right) \right\rangle &=\int d^{2}qd\omega \int d^{2}\bar{q}d%
\bar{\omega}~\left\langle \sigma \left( \mathbf{q},\omega \right) \dot{u}_{%
\mathrm{z}}\left( \mathbf{\bar{q}},\bar{\omega}\right) \right\rangle
e^{i\left( \mathbf{q}\cdot \mathbf{x}-\omega t\right) }e^{i\left( \mathbf{%
\bar{q}}\cdot \mathbf{x}-\bar{\omega}t\right) }  \notag \\
&=\int d^{2}q~S\left( q\right) P\left( q\right) \sum_{j=1}^{2}{\frac{\dot{M}%
\left( \mathbf{q},\mathbf{q}\cdot \mathbf{v}_{j}\right) C_{j}\left( \mathbf{q%
}\right) }{\left\vert M\left( \mathbf{q},\mathbf{q}\cdot \mathbf{v}%
_{j}\right) \right\vert ^{2}}}.  \label{A9}
\end{align}%
Furthermore, the correlation function $\left\langle \sigma \left( \mathbf{x}%
,t\right) u_{\mathrm{z}}\left( \mathbf{x},t\right) \right\rangle $ reads%
\begin{align}
\left\langle \sigma \left( \mathbf{x},t\right) u_{\mathrm{z}}\left( \mathbf{x%
},t\right) \right\rangle &=\int d^{2}qd\omega \int d^{2}\bar{q}d\bar{\omega}%
~\left\langle \sigma \left( \mathbf{q},\omega \right) u_{\mathrm{z}}\left( 
\mathbf{\bar{q}},\bar{\omega}\right) \right\rangle e^{i\left( \mathbf{q}%
\cdot \mathbf{x}-\omega t\right) }e^{i\left( \mathbf{\bar{q}}\cdot \mathbf{x}%
-\bar{\omega}t\right) }  \notag \\
&=\int d^{2}q~S\left( q\right) P\left( q\right) \sum_{j=1}^{2}{C_{j}\left( 
\mathbf{q}\right) \mathrm{Re}\left[ M\left( \mathbf{q},\mathbf{q}\cdot 
\mathbf{v}_{j}\right) \right] }.  \label{A10}
\end{align}%
In \ref{A9} and \ref{A10} we have added the $S\left( q\right) P\left(
q\right) $ term to take into account the partial contact in the evaluation
of $\left\langle \sigma \left( \mathbf{x},t\right) \dot{u}_{\mathrm{z}%
}\left( \mathbf{x},t\right) \right\rangle $ and $\left\langle \sigma \left( 
\mathbf{x},t\right) u_{\mathrm{z}}\left( \mathbf{x},t\right) \right\rangle $%
, where $S\left( q\right) $ has been introduced before. $M_{i}\left( \mathbf{%
q},\omega \right) =2\left[ qE_{i,\mathrm{r}}\left( \omega \right) \bar{S}%
\left( q d\right) \right] ^{-1}$ for a generic viscoelastic solid of finite
thickness, where the real function $\bar{S}(qd)$ is a correction factor
related to the adopted boundary conditions [e.g. for a slab of thickness $d$
on a rigid substrate we have%
\begin{equation*}
\bar{S}\left( qd\right) =\frac{\left( 3-4\nu \right) \cosh \left(
2qd\right) +2\left( qd\right) ^{2}-4\nu \left( 3-2\nu \right) +5}{\left(
3-4\nu \right) \sinh \left( 2qd\right) -2qd},
\end{equation*}%
where $\bar{S}\left( qd\right) \rightarrow 1$ for d$\rightarrow \infty $].

\textbf{Contact area}

The contact area $A/A_{0}=P\left( q_{1}\right) $ can be calculated from \ref%
{10}, with $G\left( \bar{q}\right) =\langle \sigma^2 \left( \mathbf{x}%
,t\right)\rangle _{\bar{q}}/\left( 2\sigma _{0}^{2}\right) $
given by \ref{A8}%
\begin{equation}
G\left( \bar{q}\right) ={\frac{1}{2\sigma _{0}^{2}}}\int_{q_{0}}^{\bar{q}%
}d^{2}q\sum_{j=1}^{2}{\frac{C_{j}\left( \mathbf{q}\right) }{\left\vert
M\left( \mathbf{q},\mathbf{q}\cdot \mathbf{v}_{j}\right) \right\vert ^{2}}}.
\label{A11}
\end{equation}

\textbf{Stored and dissipated interfacial energy}

In the process of determining the average interfacial separation $\bar{u}$
as well as the micro-rolling friction, we assume the normal approach to
occur quasi-statically with respect to the sliding kinematics. Within this
assumption (i.e. $t_{0}\gg t_{1}$, where $t_{0}\propto \dot{\bar{u}}^{-1}$
and $t_{1}\propto v^{-1}$ are, respectively, the time scale of the normal
and sliding motion), the time-derivative of the displacement field $\dot{u}_{%
\mathrm{z}}\left( \mathbf{x},t\right) $ can be expanded into two
uncorrelated sources $\dot{w}_{0}\left( \mathbf{x},t\right) +\dot{w}%
_{1}\left( \mathbf{x},t\right) $ ($\dot{w}_{1}$ is random in time), where $%
\dot{w}_{0}\left( \mathbf{x},t\right) \propto t_{0}^{-1}$ and $\dot{w}%
_{1}\left( \mathbf{x},t\right) \propto t_{1}^{-1}$. Here we calculate the
average external work over a $t_{1}$ time span as%
\begin{equation*}
W_{_{1}}=\int_{t_{1}}dt~\dot{W}=\int_{t_{1}}dt~\left[ -\sigma _{0}A_{0}\dot{%
\bar{u}}+\tau _{0}A_{0}v\right] .
\end{equation*}%
Since $t_{1}/t_{0}\ll 1$ the term $\sigma _{0}A_{0}\dot{\bar{u}}$ can be
assumed constant and given that $\int_{t_{1}}dt~\tau _{0}=\left\langle \tau
\right\rangle $, we get $W_{_{1}}=-t_{1}\sigma _{0}A_{0}\dot{\bar{u}}%
+t_{1}A_{0}v\left\langle \tau \right\rangle $. However $\dot{\bar{u}}\propto
t_{0}^{-1}$ and $v\propto t_{1}^{-1}$, resulting into%
\begin{equation*}
W_{_{1}}\approx t_{1}A_{0}v\left\langle \tau \right\rangle .
\end{equation*}%
Now the average external work over a $t_{0}$ time span reads%
\begin{equation}
W=\int_{t_{0}}dt~\dot{W}=\int_{t_{0}}dt~\left[ -\sigma _{0}A_{0}\dot{\bar{u}}%
\right] +\frac{t_{0}}{t_{1}}W_{_{1}}.  \label{A12}
\end{equation}%
However the energy conservation requires that $W=A_{0}\left\langle \sigma
\left( \mathbf{x},t\right) \dot{w}\left( \mathbf{x},t\right) \right\rangle
t_{0}=A_{0}\left\langle \sigma _{0}\left( \mathbf{x},t\right) \dot{w}%
_{0}\left( \mathbf{x},t\right) \right\rangle t_{0}+A_{0}\left\langle \sigma
_{1}\left( \mathbf{x},t\right) \dot{w}_{1}\left( \mathbf{x},t\right)
\right\rangle t_{0}$. However since $\left\langle \sigma _{0}\left( \mathbf{x%
},t\right) \dot{w}_{0}\left( \mathbf{x},t\right) \right\rangle \propto
t_{0}^{-1}$ and $\left\langle \sigma _{1}\left( \mathbf{x},t\right) \dot{w}%
_{1}\left( \mathbf{x},t\right) \right\rangle \propto t_{1}^{-1}$ we can
separate the different energy contribution in \ref{A11} as%
\begin{equation}
\left\langle \sigma _{0}\left( \mathbf{x},t\right) \dot{w}_{0}\left( \mathbf{%
x},t\right) \right\rangle t_{0}=\int_{t_{0}}dt~\left[ -\sigma _{0}\dot{\bar{u%
}}\right]  \label{A13}
\end{equation}%
\begin{equation}
A_{0}\left\langle \sigma _{1}\left( \mathbf{x},t\right) \dot{w}_{1}\left( 
\mathbf{x},t\right) \right\rangle t_{1}=W_{_{1}}.  \label{A14}
\end{equation}%
Manipulating the LHS of \ref{A13} we have%
\begin{equation*}
\left\langle \sigma _{0}\left( \mathbf{x},t\right) \dot{w}_{0}\left( \mathbf{%
x},t\right) \right\rangle t_{0}=\left\langle \sigma _{0}\left( \mathbf{x}%
,t\right) \Delta w_{0}\left( \mathbf{x},t\right) \right\rangle =\frac{1}{2}%
\Delta \left\langle \sigma _{0}\left( \mathbf{x},t\right) w_{0}\left( 
\mathbf{x},t\right) \right\rangle ,
\end{equation*}%
resulting into%
\begin{equation}
\frac{1}{2}\Delta \left\langle \sigma _{0}\left( \mathbf{x},t\right)
w_{0}\left( \mathbf{x},t\right) \right\rangle =-\sigma _{0}\Delta \bar{u},
\label{A15}
\end{equation}%
whereas \ref{A14} leads to 
\begin{equation}
\left\langle \sigma _{1}\left( \mathbf{x},t\right) \dot{w}_{1}\left( \mathbf{%
x},t\right) \right\rangle =\left( \mathbf{v}_{2}-\mathbf{v}_{1}\right) \cdot
\left\langle \mathbf{\tau }\right\rangle .  \label{A16}
\end{equation}%
Hence, by using \ref{A9} in \ref{A16} we get the sliding-induced
micro-rolling shear stress%
\begin{equation}
\left( \mathbf{v}_{2}-\mathbf{v}_{1}\right) \cdot \mathbf{\tau }\left(
p\right) =\int d^{2}q~S\left( q\right) P\left( q\right) \sum_{j=1}^{2}\frac{%
\dot{M}\left( \mathbf{q},\mathbf{q}\cdot \mathbf{v}_{j}\right) C_{j}\left( 
\mathbf{q}\right) }{\left\vert M\left( \mathbf{q},\mathbf{q}\cdot \mathbf{v}%
_{j}\right) \right\vert ^{2}},  \label{A17}
\end{equation}%
whereas the average interfacial separation can be calculated by substituting %
\ref{A10} into \ref{A15},%
\begin{equation}
\bar{u}\left( p\right) =\int d^{2}q\sum_{j=1}^{2}{C_{j}\left( \mathbf{q}%
\right) \mathrm{Re}}\left[ M\left( \mathbf{q},\mathbf{q}\cdot \mathbf{v}%
_{j}\right)^{-1} \right] \int_{p}^{\infty }d\sigma _{0}\frac{1}{2\sigma _{0}}%
\frac{\partial \left[ S\left( q,\sigma _{0}\right) P\left( q,\sigma
_{0}\right) \right] }{\partial \sigma _{0}}.  \label{A18}
\end{equation}%
For an elastic solid in contact with a hard randomly rough surface, this equation
reduces to the equation derived in Ref. \cite{Yang}.

\twocolumngrid


\begin{thebibliography}{99}
\bibitem{note} For a rectangular block if a uniform stress act on the upper
surface, the nominal stress at the interface will be uniform only if either
the Poisson ratio is zero, or else the friction at the interface vanish. If
this is not the case the normal stress will be non-uniform with
singularities at the edge of contact.
\end{thebibliography}
\end{document}